%% file: moriond2006.tex
\documentclass[11pt,a4paper]{article}
\usepackage{moriond}
\usepackage{amssymb}
\usepackage{subfigure}
\usepackage{multicol}
\usepackage{graphicx}
\usepackage{ifpdf}
\ifpdf
  \usepackage[pdftex]{hyperref}
\else
  \usepackage[ps2pdf,colorlinks=true]{hyperref}
\fi

\input{xcorr2005_macros}

\newcommand{\arcmin}{\ensuremath{{^\prime}}}

\makeatletter

\newenvironment{figurehere}
  {\def\@captype{figure}}
  {}
\makeatother

\begin{document}

\vspace*{4cm}
\title{DETECTION OF THE ISW EFFECT\\AND CORRESPONDING DARK ENERGY CONSTRAINTS}
\author{J.~D.~MCEWEN$^1$, P.~VIELVA$^2$,
    M.~P.~HOBSON$^1$, E.~MART\'{I}NEZ-GONZ\'{A}LEZ$^2$, 
    A.~N.~LASENBY$^1$}
\address{$^1$Astrophysics Group, Cavendish Laboratory, Cambridge, UK\\
         $^2$Instituto de F\'{\i}sica de Cantabria, {CSIC-Universidad de Cantabria}, Santander, Spain}
\maketitle



\newlength{\statplotwidth}
\setlength{\statplotwidth}{36mm}

\newlength{\coeffmapwidth}
\setlength{\coeffmapwidth}{53mm}

\newlength{\figspacerplot}
\setlength{\figspacerplot}{5mm}

\newlength{\figspacermap}
\setlength{\figspacermap}{3mm}

\begin{abstract}
Using a directional spherical wavelet analysis we detect the \iswtext\
(\isw) effect, indicated by a positive correlation between the
\wmaptext\ (\wmap) and \nvsstext\ (\nvss) data, at the $3.9\sigma$
level.
In a flat universe the detection of the \isw\ effect provides direct
and independent evidence for dark energy.  Moreover, we use our detection
to constrain the dark energy density \olambda.
We obtain estimates for \olambda\ consistent with other analysis techniques and data sets and rule out a zero cosmological
constant at greater than  99\% significance.
\end{abstract}

\section{Introduction}

Strong observational evidence now exists in support of the $\Lambda$ cold dark matter (\lcdm) fiducial model of the universe.
Much of this evidence comes from recent measurements of the \cmbtext\ (\cmb) anisotropies, in particular the \wmaptext\ (\wmap) data\cite{bennett:2003a}.  
At this point, the confirmation of the fiducial \lcdm\ model and the existence of dark energy by independent physical methods is of particular interest.
One such approach is through the detection of the \iswtext\ (\isw) effect\cite{sachs:1967}.

It is not feasible to separate directly the contribution of the \isw\ effect from the \cmb\ anisotropies.  Instead, as first proposed by Crittenden \&\ Turok\cite{crittenden:1996}, the \isw\ effect may be detected by cross-correlating the \cmb\ anisotropies with tracers of the local matter distribution (for redshift in the range $0\leq z \leq 2$).
A cross-correlation indicative of the \isw\ effect was detected first by Boughn \&\ Crittenden\cite{boughn:2002} and since by many other authors.  
In these proceedings we give a very brief overview of our recent work\cite{mcewen:2006} using directional spherical wavelets to detect the \isw\ effect and contrain dark energy parameters.  

\newlength{\mapwidth}
\setlength{\mapwidth}{60mm}

\newlength{\nsigmaplotwidth}
\setlength{\nsigmaplotwidth}{55mm}

\begin{minipage}{68mm}

  \begin{figurehere}
  \centering
    \subfigure[\smhw]{\includegraphics[width=\mapwidth]{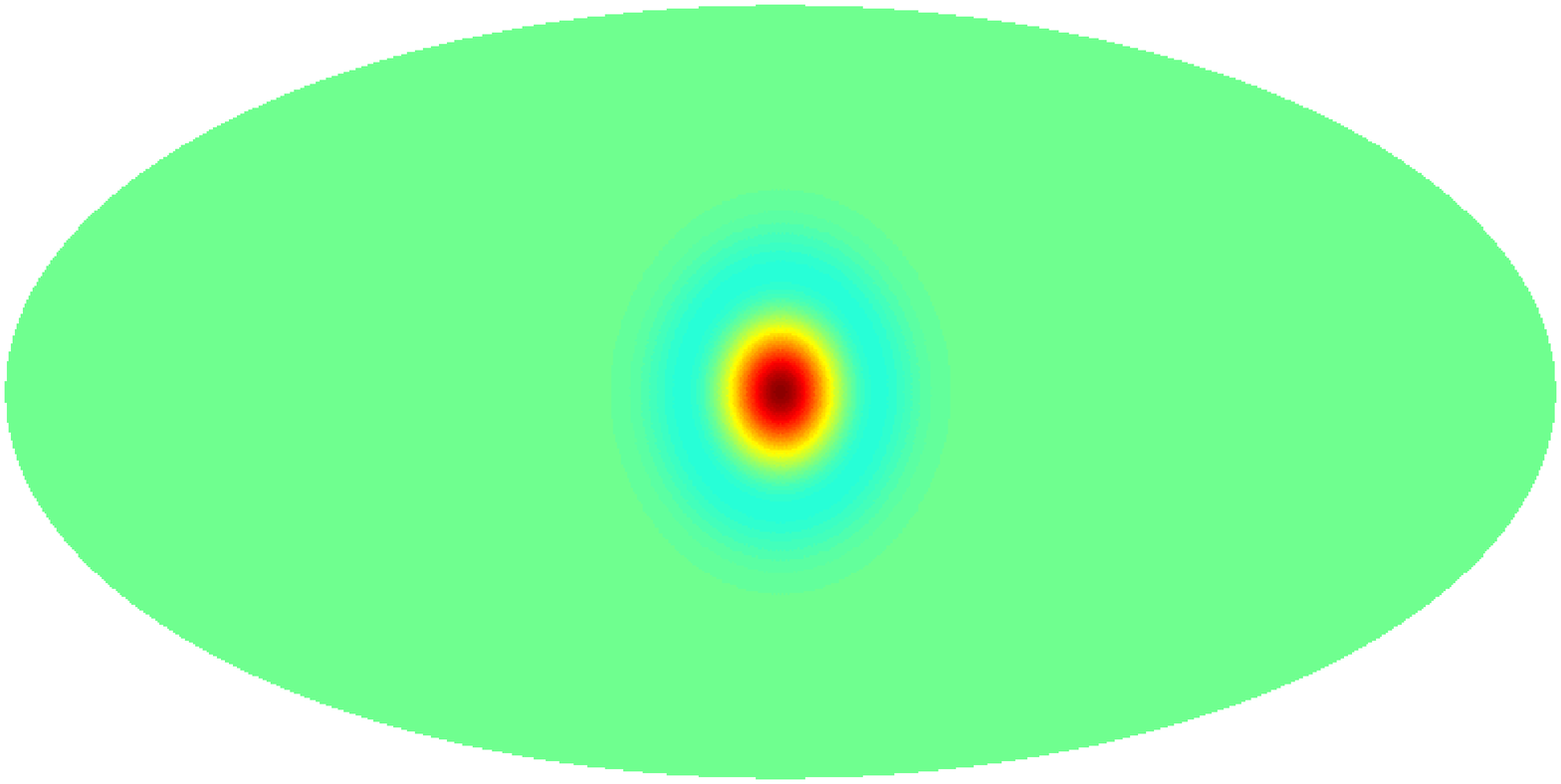}}
    \subfigure[\sbw]{\includegraphics[width=\mapwidth]{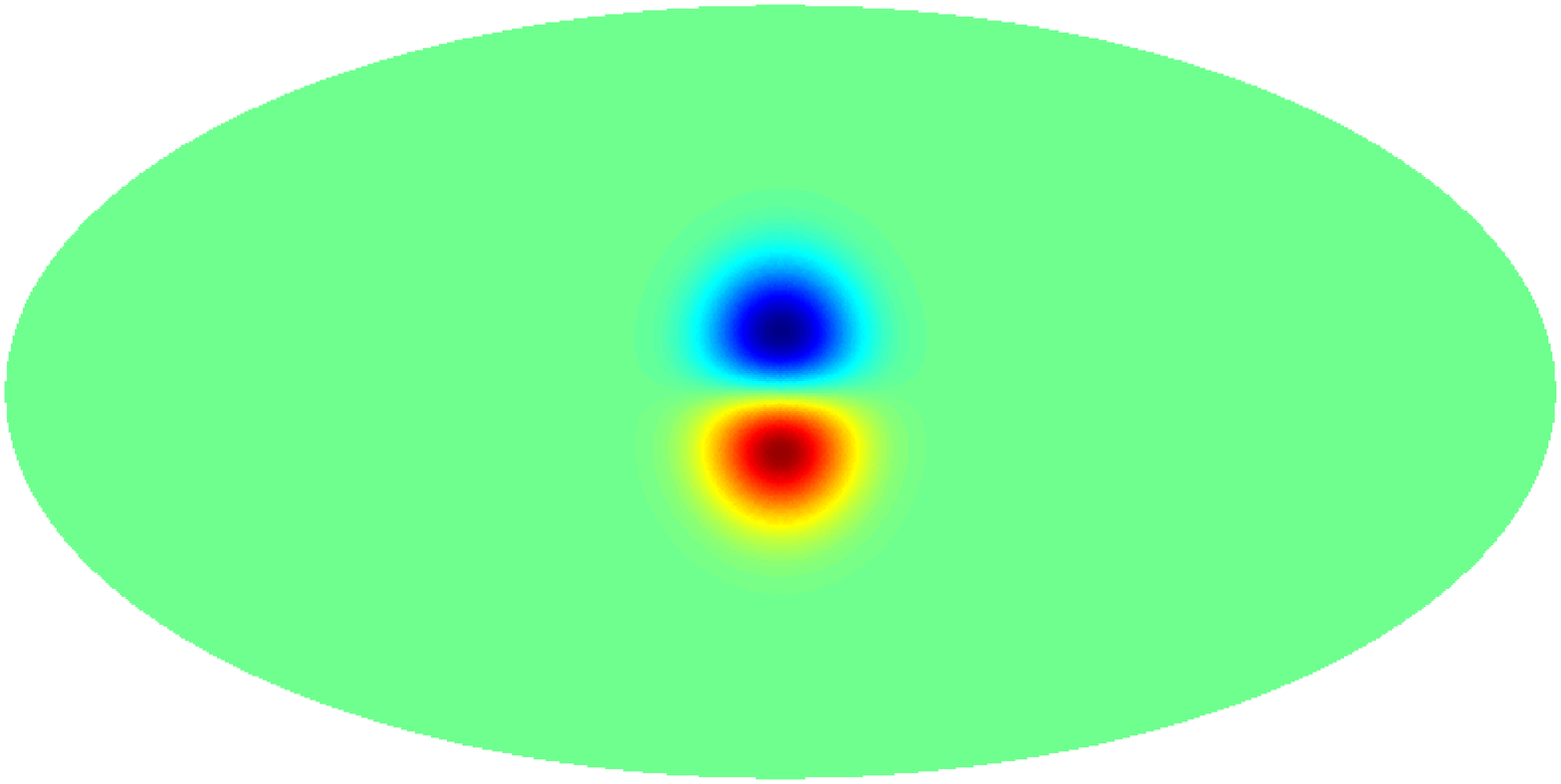}}
  \caption{Spherical wavelets at scale $\scalea=\scaleb=0.2$.}
  \label{fig:mother_wavelets}
  \end{figurehere}

\end{minipage}
\hspace{5mm}
\begin{minipage}{68mm}

  \begin{figurehere}
  \centering
  \subfigure[\smhw]
    {\includegraphics[clip=,width=\nsigmaplotwidth]{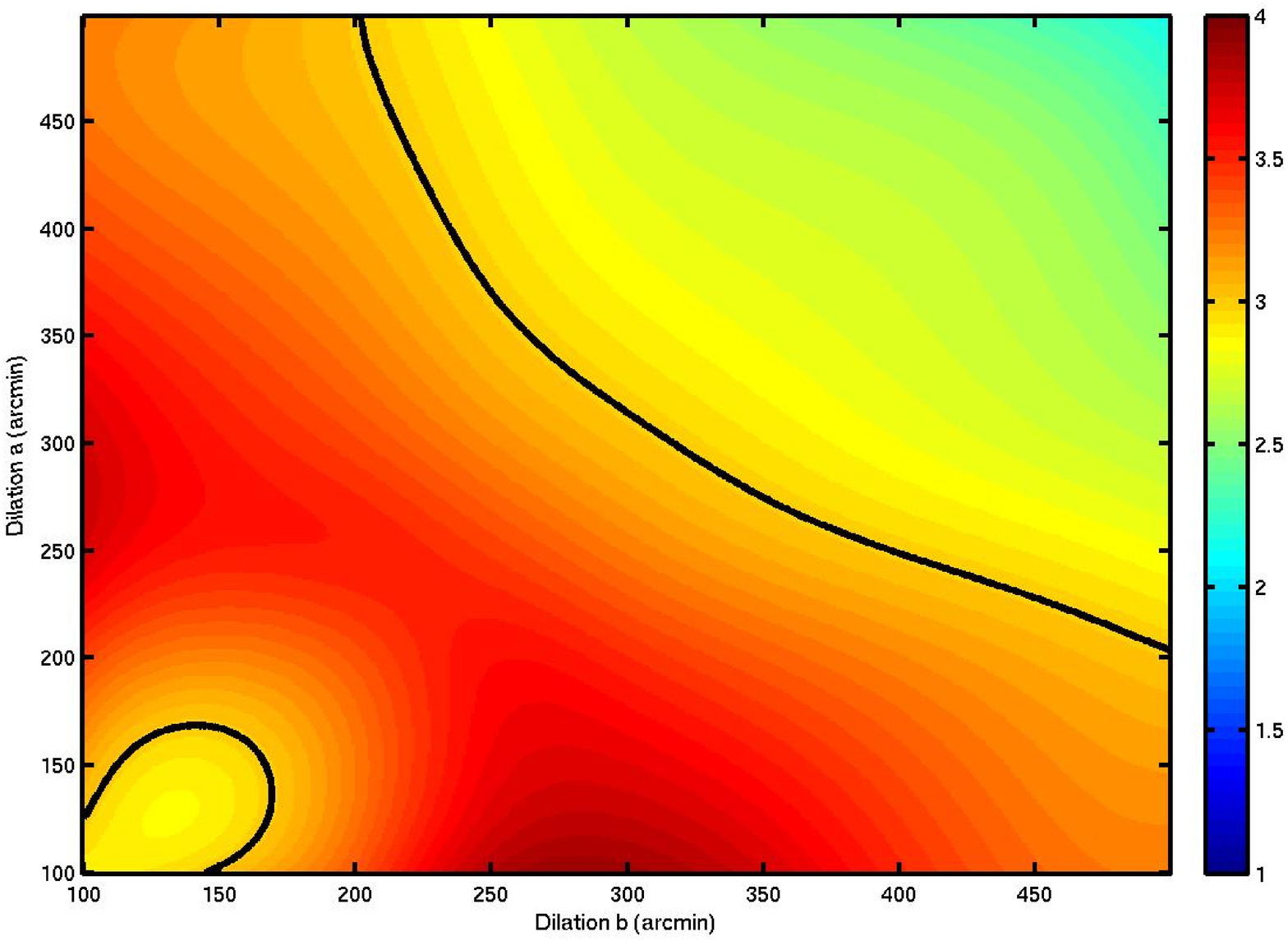}}
  \subfigure[\sbw]
    {\includegraphics[clip=,width=\nsigmaplotwidth]{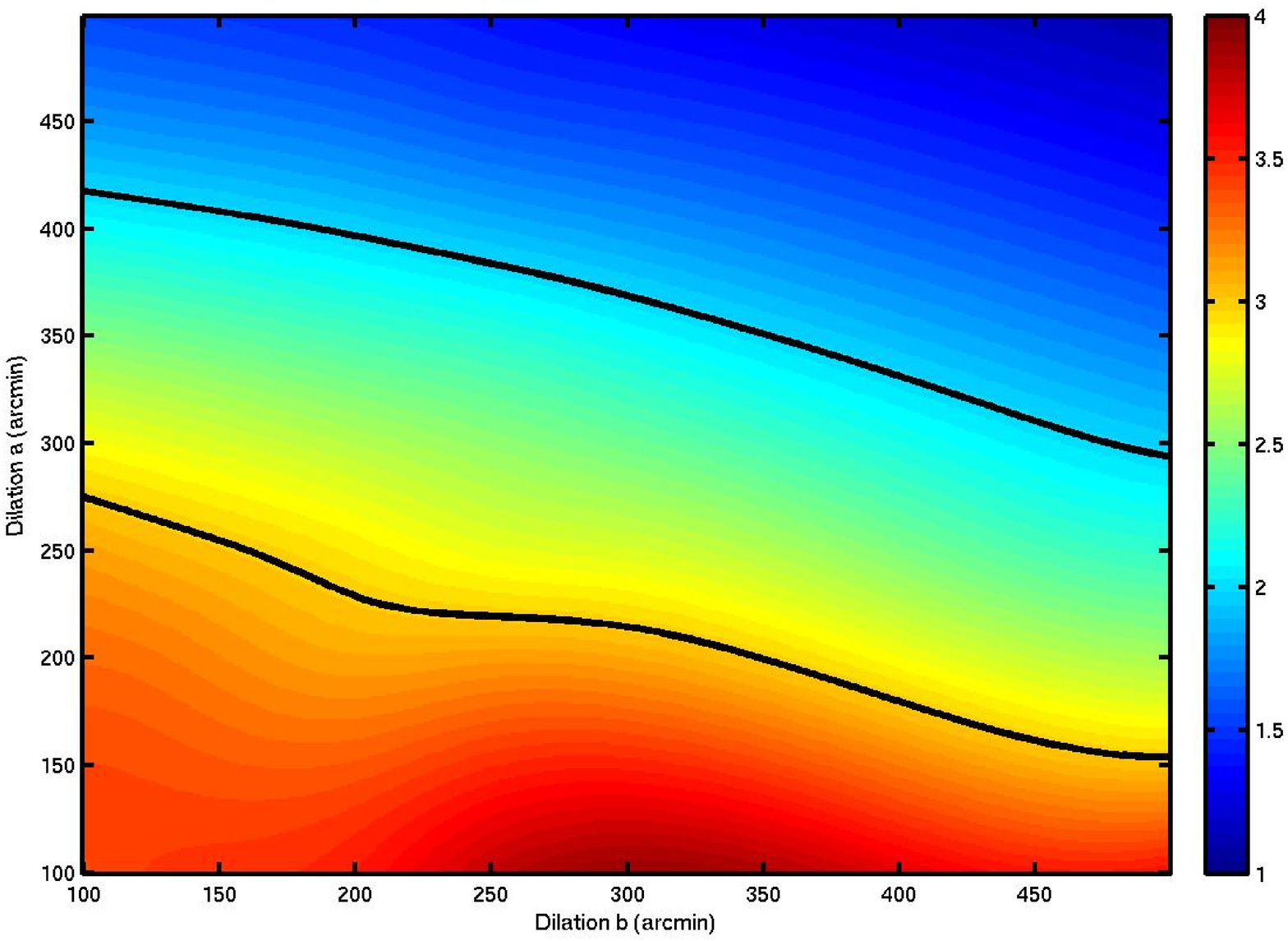}}
  \caption{Wavelet covariance \nsigma\ surfaces with two and three \nsigma\ contours shown.}
  \label{fig:nsigma}
  \end{figurehere}

\end{minipage}

\section{Analysis procedure}
\label{sec:procedure}

\subsection{\Cswttext}

Wavelets are an ideal tool to search for the \isw\ effect\cite{vielva:2006,mcewen:2006} due to the scale and spatial localisation provided by a wavelet analysis.
To perform an analysis of full-sky \cmb\ maps, Euclidean wavelets must be extended to spherical geometry.  We apply our fast \cswt\ algorithm\cite{mcewen:2005,mcewen:2004}, which is based on the spherical wavelet transform developed by Antoine, Vandergheynst and colleagues \cite{antoine:1998,wiaux:2005} and the fast spherical convolution developed by Wandelt \&\ G\'{o}rski\cite{wandelt:2001}. 
%
%
%
We consider two spherical wavelets in our subsequent analysis: the spherical Mexican hat wavelet (\smhw) and the spherical butterfly wavelet (\sbw).  These spherical wavelets are illustrated in \fig{\ref{fig:mother_wavelets}}.

\subsection{Wavelet covariance estimator}
\label{sec:wav_cov}

The covariance of the wavelet coefficients is used as an estimator to detect any cross-correlation between the \cmb\ and the local matter distribution.
The wavelet coefficient covariance estimator is denoted $\wcovest^{\ndlab\tplab}(\scaleab)$, where $(\scaleab)$ define the size of the anisotropic dilation (note that perfect reconstruction is not possible in the case of anisotropic dilations\cite{mcewen:2005}).
For a given cosmological model, the theoretical wavelet covariance is given by\cite{mcewen:2006}
\begin{equation}
\label{eqn:theoxcorr}
\wcov^{\ndlab\tplab}(\scaleab, \eulc) =
\sum_{\el=0}^\infty \:
\pixw{}^2 \:
\bm{\ndlab} \:
\bm{\tplab} \:
\clnttheo
\sum_{\m=-\el}^\el \left| (\wav_{\scaleab})_\elm \right|^2
\spcend ,
\end{equation}
where $\wav_\elm$ are the spherical harmonic coefficients of the wavelet, \clnttheo\ is the cross-power spectrum for the model considered and 
$\bm{}$ and $\pixw$ are beam and pixel window functions respectively.

\subsection{Data and simulations}

The detection of the \isw\ effect is cosmic variance limited, hence we require (near) full sky maps in our analysis.  We consider the \wmap\ co-added map and the \nvss\ radio source catalogue\cite{condon:1998}.  The near full-sky coverage and source distribution of the \nvss\ data make it a suitable probe of the local matter distribution to use.  To quantify the significance of any correlations between the data we simulate 1000 Gaussian co-added maps, constructed by mimicking the \wmap\ observing strategy and co-added map construction technique.

\subsection{Procedure}

The analysis consists of computing the wavelet covariance estimator described in \sectn{\ref{sec:wav_cov}} for a range of scales and orientations.
We consider only those scales where the \isw\ signal is expected to be significant, ranging over dilation scales from 100\arcmin--500\arcmin, in steps of 50\arcmin, and consider five evenly spaced orientations in the domain $[0,\pi)$.
Any deviation from zero in the wavelet covariance estimator for any particular scale or orientation is an indication of a correlation between the \wmap\ and \nvss\ data and hence a possible detection of the \isw\ effect.  An identical analysis is performed using the simulated co-added \cmb\ maps in order to construct significance measures for any detections made.
Finally, we use any detections of the \isw\ effect to constrain dark energy parameters.

\section{Results and discussion}
\label{sec:results}

\subsection{Detection of the \isw\ effect}

A positive wavelet covariance outside of the 99\% significance level is detected on a number of scales and orientations.  On examining the distribution of the wavelet covariance statistics from the simulations, the covariance statistics appear to be approximately Gaussian distributed.  This implies that the approximate significance of any detections of a non-zero covariance can be inferred directly from the \nsigma\ level.  In \fig{\ref{fig:nsigma}} we plot the \nsigma\ surfaces for each wavelet in $(\scaleab)$ space.  
The maximum detection made for each wavelet occurs at $\nsigma=3.9$ on wavelet scales about $(\scalea,\scaleb)=(100\arcmin,300\arcmin)$.
The wavelet analysis allows us to localise on the sky those regions that contribute most strongly to the covariance detected.  These localised regions are evenly distributed over the entire sky and do not appear to be the sole source of the correlation between the data.\cite{mcewen:2006}
In addition, we test whether foregrounds or \wmap\ systematics are responsible for the correlation but find no evidence to support this.\cite{mcewen:2006}  
The correlation we detect therefore appears to be consistent with a signal due to the \isw\ effect.

%

\subsection{Constraints on dark energy}

We use our detection of the \isw\ effect to constrain dark energy parameters by comparing the theoretically predicted wavelet covariance signal for different cosmological models with that measured from the data.  In particular, we constrain  \olambda\ over the range $0<\olambda<0.95$, assuming a pure cosmological constant (we allow the equation-of-state parameter \w\ to vary in our previous work\cite{mcewen:2006}).  For other cosmological parameters we assume concordance model values\cite{spergel:2003}. 
The likelihood distributions computed using each wavelet are shown in \fig{\ref{fig:pdfwone}}.  
Within error bounds the parameter estimates obtained from the mean of the distributions are consistent with each other and with estimates made using other techniques and data sets.
We also show in \fig{\ref{fig:pdfwone}} the cumulative probability \mbox{$P(\olambda>x)$}.  For both wavelets, we have very strong evidence for the existence of dark energy.

\newlength{\pdfwidth}
\setlength{\pdfwidth}{65mm}

\begin{figure*}
\centering
\begin{minipage}{\textwidth}
\begin{multicols}{2}
  \centering
  \subfigure[\smhw\ likelihood distribution for \olambda]{\includegraphics[width=\pdfwidth]{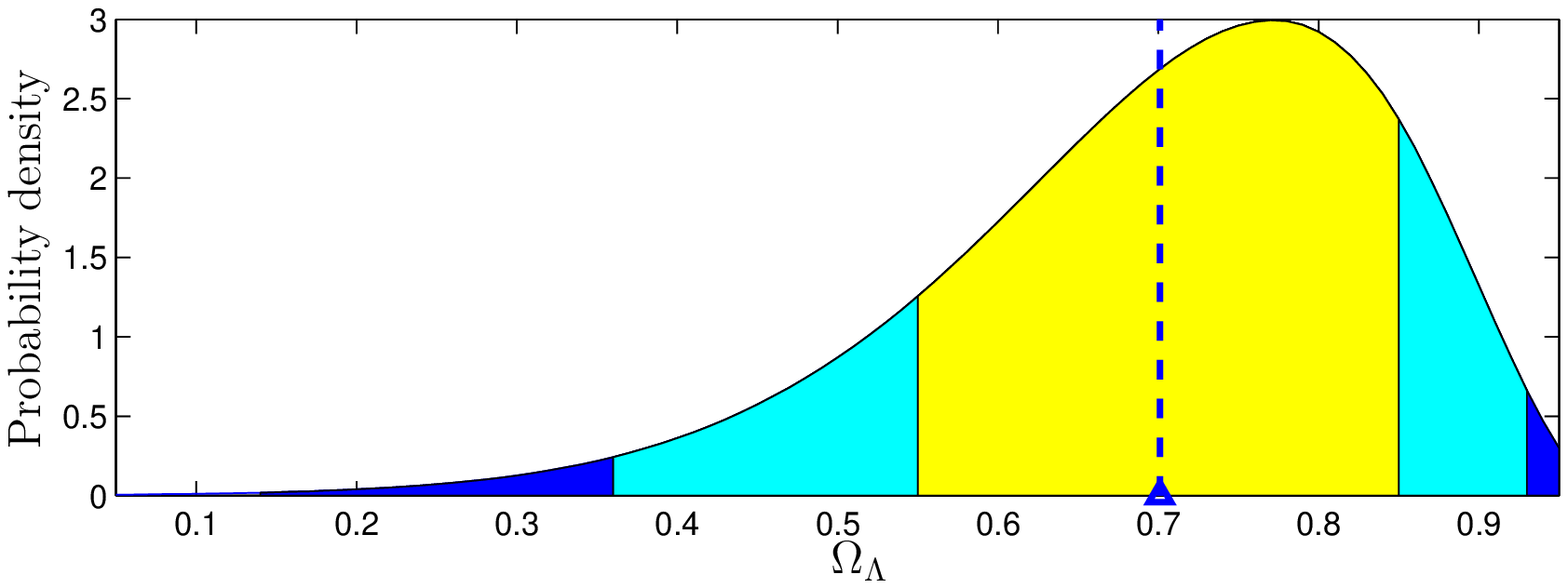}} \\
  \subfigure[\sbw\ likelihood distribution for \olambda]{\includegraphics[width=\pdfwidth]{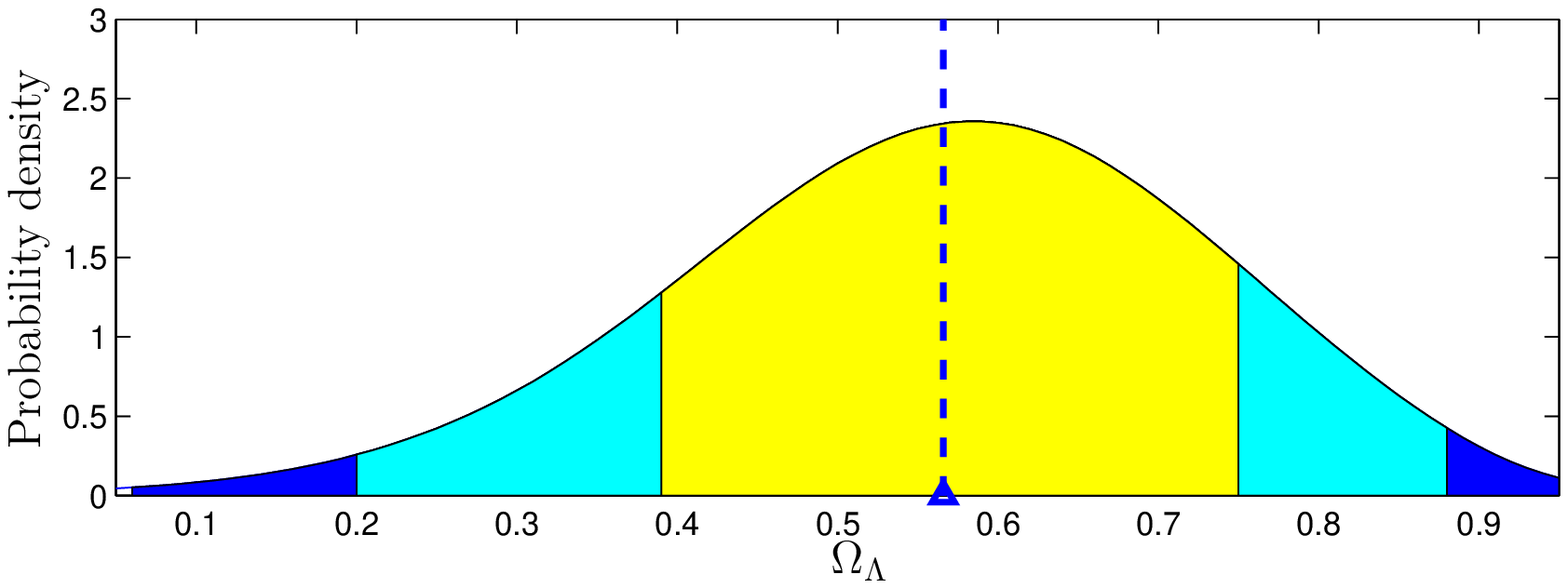}} 
  \newpage
  
  \subfigure[Cumulative probability functions $P(\olambda>x)$ ]{\includegraphics[width=72mm]{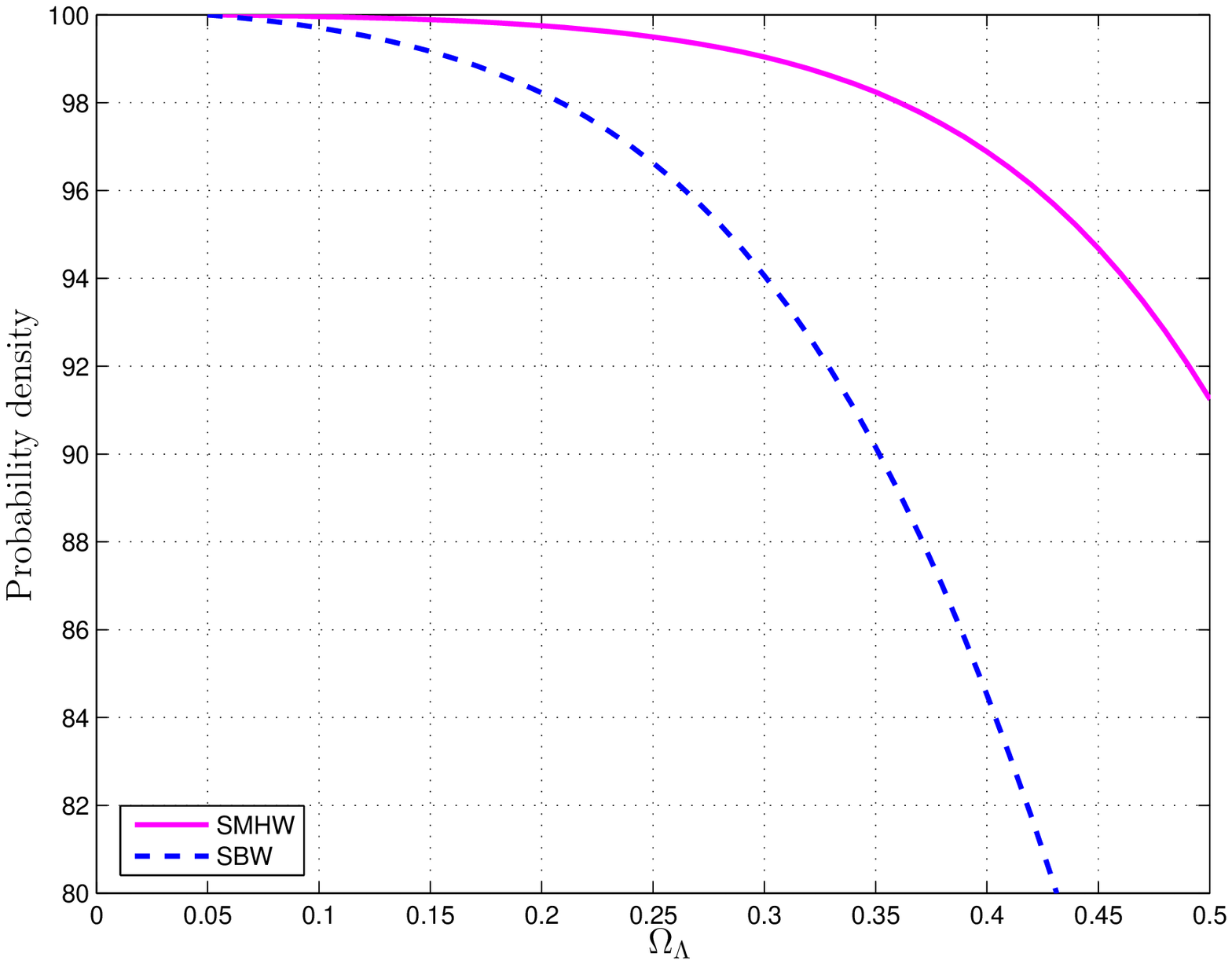}} 
\end{multicols}
\end{minipage}
\caption{Likelihood distributions and cumulative probability functions for \olambda\ when $\w=-1$.
Confidence regions at 68\% (yellow), 95\% (light-blue) and 99\% (dark-blue) are also shown on the likelihood distributions, with the parameter estimates made from the mean of the distribution shown by the triangle and dashed line.  The cummulative probability functions show the probability $P(\olambda>x)$ for the \smhw\ (solid) and \sbw\ (dashed).
}
\label{fig:pdfwone}
\end{figure*}

\section{Conclusions}
\label{sec:conclusions}

Using directional spherical wavelets we have made a detection of the \isw\ effect at the $3.9\sigma$ level.  
In a flat universe the \isw\ effect exists only in the presence of dark energy, hence our detection may be interpreted as independent evidence for dark energy.
We rule out a model with no dark energy ($\olambda=0$) at $>99$\% significance.

\section*{Acknowledgements}

We thank Jacques Delabrouille and Daniel Mortlock for insightful discussions and comments.  See McEwen \etal\ 2006\cite{mcewen:2006} for additional acknowledgements.

\section*{References}



\end{document}

%% file: xcorr2005_macros.tex
\newcommand{\fig}[1]{Fig.~#1}

\newcommand{\sectn}[1]{section~#1}

\newcommand{\etal}{\mbox{et al.}}

\newcommand{\mnras}{Mon.\ Not.\ Roy.\ Astron.\ Soc.}
\newcommand{\physd}{\mbox{Phys.\ Rev.\ D.}}
\newcommand{\physlett}{\mbox{Phys.\ Rev.\ Lett.}}

\newcommand{\apj}{ApJ}

\newcommand{\apjs}{ApJS}

\newcommand{\jmp}{J.\ of Math.\ Phys.}

\newcommand{\cmb}{{CMB}}
\newcommand{\cmbtext}{{cosmic microwave background}}

\newcommand{\wmap}{{WMAP}}
\newcommand{\wmaptext}{{Wilkinson Microwave Anisotropy Probe}}

\newcommand{\isw}{{ISW}}
\newcommand{\iswtext}{{integrated Sachs-Wolfe}}

\newcommand{\cswt}{{CSWT}}

\newcommand{\Cswttext}{Continuous spherical wavelet transform}

\newcommand{\nvss}{{NVSS}}
\newcommand{\nvsstext}{{NRAO VLA Sky Survey}}

\newcommand{\lcdm}{\ensuremath{\Lambda}{CDM}}

\newcommand{\smhw}{{\rm SMHW}}

\newcommand{\sbw}{{\rm SBW}}

\newcommand{\olambda}{\ensuremath{\Omega_\Lambda}}
\newcommand{\w}{\ensuremath{w}}

\newcommand{\clnttheo}{\ensuremath{C_\el^{\rm NT}}}

\newcommand{\ndlab}{\ensuremath{{\rm N}}}
\newcommand{\tplab}{\ensuremath{{\rm T}}}

\newcommand{\wcov}{\ensuremath{X_\wav}}
\newcommand{\wcovest}{\ensuremath{\hat{X}_\wav}}

\newcommand{\pixw}{\ensuremath{p_\el}}
\newcommand{\bm}[1]{\ensuremath{b_\el^{#1}}}

\newcommand{\el}{\ensuremath{\ell}}
\newcommand{\m}{\ensuremath{m}}
\newcommand{\elm}{\ensuremath{{\el\m}}}

\newcommand{\spcend}{\ensuremath{\:}}

\newcommand{\eulc}{\ensuremath{\gamma}}

\newcommand{\scalea}{\ensuremath{a}}
\newcommand{\scaleb}{\ensuremath{b}}
\newcommand{\scaleab}{\ensuremath{{\scalea,\scaleb}}}

\newcommand{\wav}{\ensuremath{\psi}}

\newcommand{\nsigma}{\ensuremath{N_\sigma}}

%% file: moriond2006.bbl
\begin{thebibliography}{99}




  \bibitem{antoine:1998}
    Antoine J. -P. and Vandergheynst P., 1998,
    \jmp, 39, 8, 3987






  \bibitem{bennett:2003a}
     Bennett C. L. \etal\ (\wmap\ team), 2003a, \apjs, 148, 1





  \bibitem{boughn:2002}
    Boughn S. P., Crittenden R. G., 2002, \physlett, 88, 21302

%


%
%
%
%
%

  \bibitem{condon:1998}
    Condon J. J., 1998, \apj, 115, 1693


  \bibitem{crittenden:1996}
    Crittenden R. G., Turok N., 1996, \physlett, 76, 575




%


%
%
%
 






%


%
%
%
%
%
%
%
%
%


  \bibitem{mcewen:2006}
    McEwen J. D., Vielva P., Hobson M. P., Mart\'{\i}nez-Gonz\'{a}lez E., Lasenby A. N., 2006,
    submitted to \mnras\ (astro-ph/0602398) 

  \bibitem{mcewen:2005}
    McEwen J. D., Hobson M. P., Mortlock D. J., Lasenby A. N., 2005,
    submitted to IEEE Trans.\ Sig.\ Proc.\ (astro-ph/0506308)

  \bibitem{mcewen:2004}
    McEwen J. D., Hobson M. P., Lasenby A. N., Mortlock D. J., 2004,
    XXXIXth Recontres de Moriond

%



%







%


%
  \bibitem{sachs:1967}
    Sachs R. K., Wolfe A. M., 1967, \apj, 147, 73



%

  \bibitem{spergel:2003}
     Spergel D. N. \etal\ (\wmap\ team), 2003, \apjs, 148, 175


%

  \bibitem{vielva:2006}
    Vielva P., Mart\'{\i}nez-Gonz\'{a}lez E., Tucci M., 2006,
    \mnras, 365, 891 


   \bibitem{wandelt:2001}
     Wandelt B. D., G\'{o}rski K. M., 2001,
     \physd, 63, 123002, 1
%


  \bibitem{wiaux:2005} Wiaux Y., Jacques L., Vandergheynst P.,
    2005, \apj, 632, 15



\end{thebibliography}
